\documentstyle[amssymb,twocolumn,aps,prl,epsfig]{revtex}

\twocolumn

\begin{document}
\draft
\title{Comparative study of the transient evolution of Hanle EIT/EIA resonances. }
\author{P. Valente, H. Failache and A. Lezama\thanks{%
E-mail: alezama@fing.edu.uy}}
\address{Instituto de F\'{i}sica, Facultad de Ingenier\'{i}a. Casilla de correo 30. \\
11000, Montevideo, Uruguay.}
\date{\today }
\maketitle

\begin{abstract}
The temporal evolutions of coherent resonances corresponding to
electromagnetically induced transparency (EIT) and absorption (EIA) \ were
observed in a Hanle absorption experiment carried on the $D_{2}$ lines of $%
^{87}$Rb vapor by suddenly turning the magnetic field on or off. The main
features of the experimental observations are well reproduced by a
theoretical model based on Bloch equation where the atomic level degeneracy
has been fully accounted for. Similar (opposite phase) evolutions were
observed at low optical field intensities for Hanle/EIT or Hanle/EIA
resonances. Unlike the Hanle/EIA\ transients which are increasingly shorter
for driving field intensities approaching saturation, the $B\neq 0$
transient of the Hanle/EIT signal at large driving field intensities present
a long decay time approaching the atomic transit time. Such counterintuitive
behavior is interpreted as a consequence of the Zeno effect.
\end{abstract}


\preprint{}

\section{Introduction.}

Large attention has been paid in recent years to the fascinating properties
of macroscopic samples of atoms or molecules prepared in a specific linear
combination of quantum states \cite{SCULLYREP}. Such media are said to be
coherently prepared and its statistical (macroscopic) description
corresponds to a density matrix with nonzero off-diagonal (coherence) terms.
A wide variety of physical consequences of coherently prepared media have
been observed and many crucial applications achieved \cite{SCULLYBOOK}. A
few examples are: Coherence population trapping (CPT) \cite
{ALZETTA76,ARIMONDOREV} successfully exploited for sub-recoil laser cooling 
\cite{ASPECT89}. Electromagnetically induced transparency (EIT) and the
related topic of laser without inversion \cite{HARRIS97}. Enhancement of
optical nonlinearities \cite{HARRIS90} and efficient frequency generation 
\cite{MERRIAM00}. Very large dispersion \cite{SCULLY92} and its application
to sensitive magnetometry \cite{NAGEL98} and optical propagation with slow
group velocity \cite{HAU99,KASH99}.

Most experiments and theoretical modelling on CPT, EIT and related coherence
effects deal with three level systems in a $\Lambda $ configuration where
the two long-living lower levels have different energies. This requires the
consideration of two distinct optical fields (coupling and probe fields)
acting on either arms of the $\Lambda $ system. However, interesting energy
level configurations for coherent spectroscopy purposes can also be obtained
using the Zeeman sublevels of a degenerate two-level atomic transition.

The coherent spectroscopy of degenerate two-level systems (DTLS) was
recently explored using mutually coherent optical fields \cite
{AKULSHIN98,LEZAMA99,AKULSHIN99,LEZAMA00}. A new coherent effect emerged
corresponding to a resonant increase of the atomic absorption. Such effect
was designated electromagnetically induced absorption (EIA). EIA is observed
in DTLS in closed transitions with a higher degeneracy in the upper level 
\cite{AKULSHIN98,LEZAMA99}. Recently, coherent resonances in DTLS were used
to produce very slow positive and negative \cite{AKULSHIN99} group velocity
and to obtain ``light storage'' \cite{PHILLIPS01}.

A major advantage of the use of DTLS is that a single optical field may be
enough to induce coherent effects. Indeed, the two orthogonal polarization
components ($\sigma ^{+}$ and $\sigma ^{-}$) of the same linearly polarized
optical wave couple different Zeeman sublevels of the ground and excited
state. The Raman resonance condition between ground state sublevels is then
automatically reached at perfect ground state degeneracy or destroyed by the
application of a static magnetic field. As a result, the atomic response of
a DTLS to the excitation by a single optical field with linear polarization
presents sharp variations as a function of the magnetic field around zero
magnetic field. This is in essence the well known (ground state) Hanle
effect \cite{HANLE24,KASTLER73,CORNEY77} intimately connected to the Zeeman
optical pumping \cite{HAPPER72}.

CPT and Hanle effects were theoretically and experimentally analyzed within
a common frame by Renzoni and coworkers \cite{RENZONI97}. They studied the
hyperfine (open) transitions of the $D_1$ line of Na. This work was followed
by theoretical investigation of Hanle/CPT resonances in open transitions 
\cite{RENZONI98,RENZONI99,RENZONI00}. Hanle/CPT resonances were also studied
on the $D_1$ and $D_2$ lines of Rb in a vapor cell experiment. Inverted
Hanle resonances (increased absorption) were then observed in the case of $%
F_g\rightarrow F_e=F_g+1$ transitions \cite{DANCHEVA00}. These resonances
are related to the EIA effect previously reported in \cite
{AKULSHIN98,LEZAMA99}. A theoretical investigation of the enhanced
absorption Hanle resonances has recently been presented in \cite{RENZONI01}.

The temporal evolution of EIT signals was theoretically analyzed by Li and
coworkers \cite{LI95} and experimentally investigated by Chen {\it et al}. 
\cite{CHEN98} for strong coupling field intensity. The low driving field
intensity case was initially considered in \cite{JYOTSNA95}. The influence
of several relaxation mechanisms such as time-of-flight, dephasing
collisions and velocity changing collisions was studied in \cite{ARIMONDO96}%
. A refined treatment of the influence of the time-of-flight on the
transient atomic response in coherence resonances was discussed in \cite
{RENZONI98,RENZONI99}.

This paper is concerned with the study of the temporal evolution of the
Hanle signal as the Raman resonance condition between the optical field and
the ground state Zeeman sublevels is suddenly achieved or destroyed by
turning off or on a static longitudinal magnetic field. We focus on the
comparison between the temporal evolution of the Hanle/EIT (reduced
absorption) and Hanle/EIA\ (increased absorption) resonances at various
driving field intensities. The experiments were carried on the $D_{2}$ lines
of Rb in a vapor cell. Although we studied both stable Rb isotopes, only the
results concerning $^{87}$Rb will be presented here. The two cases
(Hanle/EIT or Hanle/EIA resonances) can be observed depending on which
ground state hyperfine level is excited. As already pointed out \cite
{AKULSHIN98,LEZAMA99,DANCHEVA00,RENZONI01}, the atomic response when the
lower alkaline atom ground state hyperfine level is excited corresponds, at
the Raman resonance, to increased transparency (EIT). Conversely, the
response obtained when the upper ground state hyperfine level is excited
results in an inverted Hanle resonance corresponding to an increase of the
atomic absorption (EIA). For an EIT type transition, the transient \
following the cancellation of the magnetic field can be interpreted as the
falling of the atomic system into the uncoupled or dark state (DS). For an
EIA type transition, the transient corresponds to the atomic system evolving
towards the enhanced absorption state (EAS) \cite{RENZONI01}. We have also
studied the transients occurring when the Raman resonance condition is
suddenly modified by the application of a magnetic field producing a Zeeman
shift of the ground state sublevels which is larger than the coherence
resonance width at low light intensity but smaller than the excited state
width. This transient evolution correspond to the system leaving the DS or
the EAS in the case of EIT or EIA type transitions respectively.

The experiments are described in the next section of the paper. Section
three is devoted to the discussion of the experimental results in view of a
theoretical model of the atomic evolution. Section four presents the
conclusions of this work.

\section{Experiment.}

\begin{figure}[tbp]
\begin{center}
\mbox{\epsfig{file=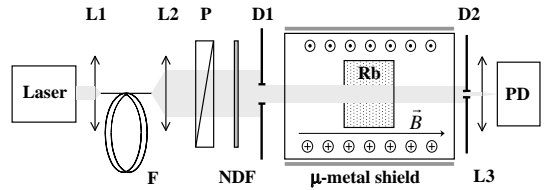,width=3.5in}}
\end{center}
\caption{Scheme of the experimental setup (L1, L2, L3: lenses, F: mono-mode
optical fiber, P: polarizer, NDF: neutral density filter, D1, D2: iris
diaphragms, PD: photodiode).}
\label{setup}
\end{figure}

The experimental setup scheme is shown in Fig. \ref{setup}. A $2\ cm$ long
glass cell containing a mixture of $^{85}$Rb and $^{87}$Rb vapor and no
buffer gas was used. The cell was slightly heated above room temperature to
obtain around $70\%$ resonant absorption. The cell was placed inside a
cylindrical coil for magnetic field control. The coil and the cell were
placed inside a cylindrical $\mu $-metal shield to reduce magnetic fields
components perpendicular to the cylindrical coil axis to less than $10\ mG$.
The atomic sample was illuminated with a $1\ mW$ laser beam issued from an
injection locked diode laser (linewidth $<1\ MHz$) whose frequency could be
tuned and stabilized along the Rb $D_{2}$ lines ($780\ nm$). The precise
laser frequency position was monitored with respect to an auxiliary
saturated absorption setup. The laser light was spatially filtered using a $%
50\ cm$ long single mode optical fiber. The linearly polarized laser
propagated along the direction of the magnetic coil axis. The intensity of
the light was controlled with neutral density filters. An iris diaphragm
placed before the cell defined the beam cross section at the atomic sample.
A second diaphragm, with smaller diameter, placed after the cell selects the
central part of the transmitted beam. The transmission was monitored with an
avalanche photodiode ($100\ MHz$ bandwidth) and recorded in a digitizing
oscilloscope.

To study the transient behavior of the Hanle/EIT(/EIA) resonances the
longitudinal magnetic field was periodically switched between two different
constant values $B_{0}=0$ and $B_{1}\simeq 250\ mG$ while observing the
temporal variation of the transmitted light power. For this, the coil
current was driven by the square-wave output of a signal generator at $10\
kHz$. After being switched on or off the magnetic field reached a new
stationary value in approximately $0.5\ \mu s.$ During the recording of the
absorption transients, the iris diaphragm placed after the cell had a
diameter at least a factor of two smaller than the diameter of the iris
placed before the cell and defining the beam cross section. By this means,
one ensures that the collected light originates from atoms whose transverse
path across the cylindrical light beam is close to a diameter. For these
atoms the mean transverse transit time across the beam can be estimated as $%
\tau =D\left( 2k_{B}T/m\right) ^{-1/2}$ where $D$ is the beam diameter, $T$
the vapor temperature, $k_{B}$ the Boltzman constant and $m$ the atom mass
(For Rb $\tau \simeq 40$\ $\mu s$ for $D=1\ cm$ \ and $T=330\ K$).

\begin{figure}[tbp]
\begin{center}
\mbox{\epsfig{file=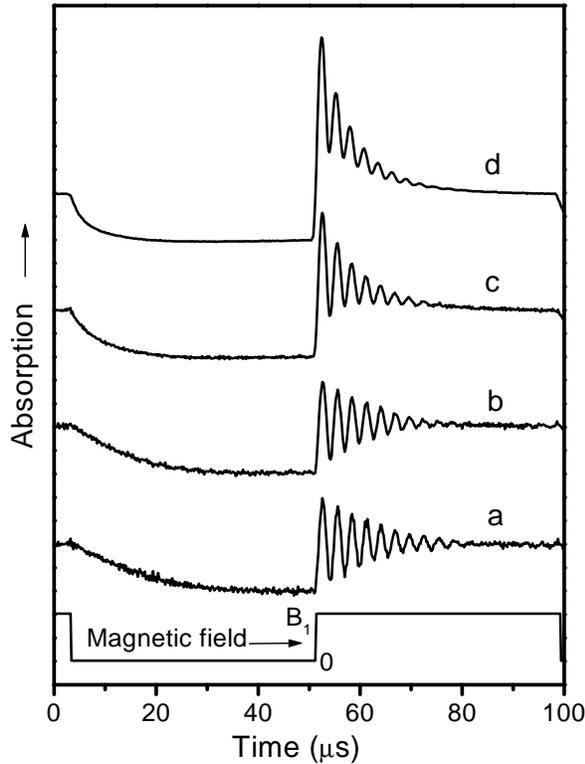,width=3.5in}}
\end{center}
\caption{Observed transient evolution of the Hanle/EIT resonance in $^{87}Rb$
for different driving field intensities $I$. a) $I\simeq 30\ \protect\mu
W/cm^{2}$, b) $I\simeq 90\ \ \protect\mu W/cm^{2}$, c) $I\simeq 0.3\%
mW/cm^{2}$, d) $I\simeq 0.9\ mW/cm^{2}$. Magnetic field $B_{1}\simeq 250\ mG$%
. Driving field tuned near the peak of the $5S_{1/2}\left( F=1\right)
\rightarrow 5P_{3/2}$ Doppler broadened absorption line.}
\label{intensdepeit}
\end{figure}

\begin{figure}[tbp]
\begin{center}
\mbox{\epsfig{file=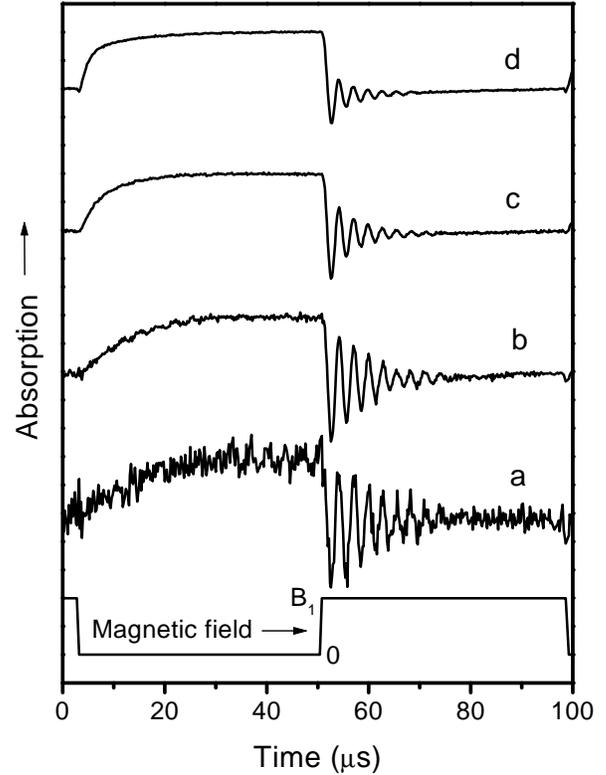,width=3.5in}}
\end{center}
\caption{Observed transient evolution of the Hanle/EIA resonance in $^{87}Rb$
for different driving field intensities $I$. a) $I\simeq 30\ \protect\mu
W/cm^{2}$, b) $I\simeq 90\ \ \protect\mu W/cm^{2}$, c) $I\simeq 0.3\%
mW/cm^{2}$, d) $I\simeq 0.9\ mW/cm^{2}$. Magnetic field $B_{1}\simeq 250\ mG$%
. Driving field tuned near the peak $5S_{1/2}\left( F=2\right) \rightarrow
5P_{3/2}$ Doppler broadened absorption line.}
\label{intensdepeia}
\end{figure}

The transient absorption records obtained for EIT and EIA type transitions
are shown in Figs. \ref{intensdepeit} and \ref{intensdepeia} respectively.
When the magnetic field is switched off, the atomic system evolves
exponentially towards a new steady state which corresponds to the DS in the
case of an EIT type transition or the EAS in the case of an EIA type
transition. A quite different behavior is observed when the magnetic field
is suddenly restored. The evolution towards the new steady state is then a
damped oscillation at a frequency given by twice the ground state Zeeman
frequency produced by the magnetic field. This oscillation corresponds to
the Larmor precession of the coherently prepared DS or EAS in the presence
of the static magnetic field. Notice the opposite phase corresponding to the
EIT and EIA type transitions.

\begin{figure}[tbp]
\begin{center}
\mbox{\epsfig{file=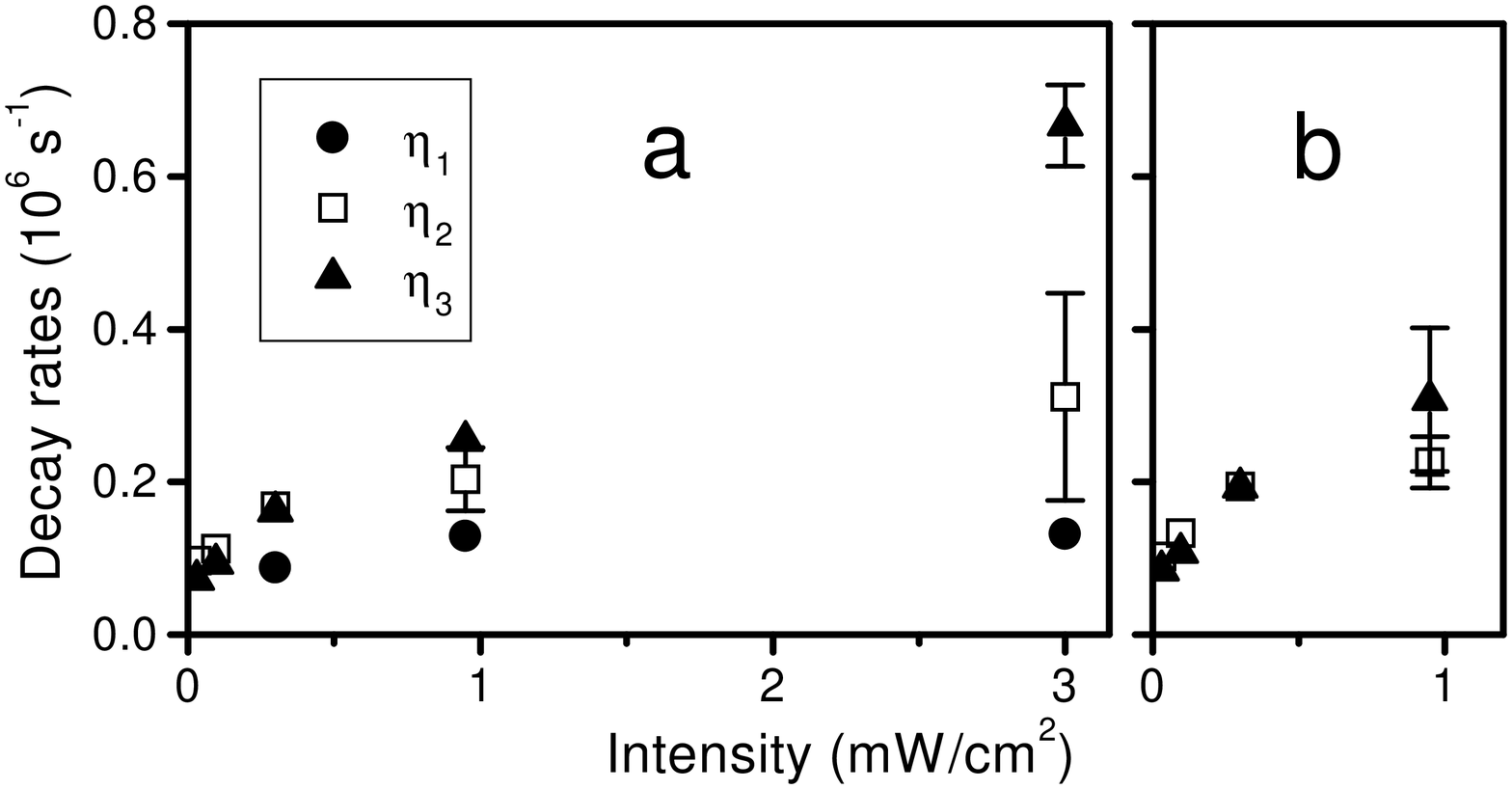,width=3.5in}}
\end{center}
\caption{Intensity dependence of the exponential decay rate $\protect\eta %
_{2}$ observed for $B=0$, \ the exponential decay rate $\protect\eta _{1}$
of the non oscillating component and the damping rate $\protect\eta _{3}$ of
the oscillating component of the transients observed for $B\neq 0$. a) EIT
type transition. b) EIA type transition. Light beam diameter at the cell $%
13\ mm$.}
\label{decays}
\end{figure}

Figs. \ref{intensdepeit} and \ref{intensdepeia} present the variations with
light intensity of the transient evolutions of EIT and EIA type transitions
respectively. In these series, the laser beam diameter is kept fixed at $%
1.3\ cm$ while its intensity is varied with neutral density filters. The
plots have been rescaled in order to present the same difference between the
two steady states regimes. At all intensities, the transient evolution
towards the steady state regime with $B=0$ is well described by an
exponential decay. Similar decay times are observed at $B=0$ for a given
laser intensity for EIT and EIA\ type transitions. This is still the case of
the $B\neq 0$ transients at low intensities, the damping of the oscillating
transients occur with similar characteristic decay rates for the EIT and
EIA\ type transitions. These rates, which are also comparable to the $B=0$
exponential decay rates observed for the same light intensity are of same
magnitude than the inverse of the estimated time-of-flight across the light
beam. Significant differences between the EIT and EIA transients arise for\ $%
B\neq 0$ as the light intensity is increased. In the EIA type transition
(Fig. \ref{intensdepeia}) increasing the light intensity results in a faster
damping of the oscillating transient. The damping rate of the oscillation
closely follows the exponential decay rate of the corresponding $B=0$
transient. The behavior is rather different for the $B\neq 0$ transient in
the EIT type transition. In this case the temporal evolution significantly
deviates from a single sine-damped oscillation. It is better described by
the sum of a sine-damped oscillation plus a non oscillating exponentially
decaying term. The characteristic decay rate of this non oscillating term is
rather insensitive to the laser intensity and remains comparable to the
time-of-flight decay rate. For a quantitative analysis, the observed
temporal evolutions were (least square) fitted with the function $%
y_{1}\left( t\right) =Aexp\left( -\eta _{2}t\right) $ in the case of the $%
B=0 $ transients and with the function $y_{2}\left( t\right) =Cexp\left(
-\eta _{1}t\right) +Dexp\left( -\eta _{3}t\right) sin(\beta t+\varphi )$ for
the $B\neq 0$ transients ($A,C,D,\eta _{1},\eta _{2},\eta _{3},\beta
,\varphi $ are adjustable parameters). Since the $B\neq 0$ transient for the
EIA type transition does not show any significant non oscillating term, in
this case, coefficient $C$ was taken zero. The fitting functions closely
adjust to the data (the differences would be barely observable on the scale
of Figs. \ref{intensdepeit} and \ref{intensdepeia}). The fitted values of
the decay rates and the oscillation frequency are plotted in Fig. \ref
{decays}. Notice that $\eta _{0}$ and $\eta _{2}$ are growing functions of
the laser intensity $I$ while $\eta _{1}$ remains approximately constant.

\begin{figure}[tbp]
\begin{center}
\mbox{\epsfig{file=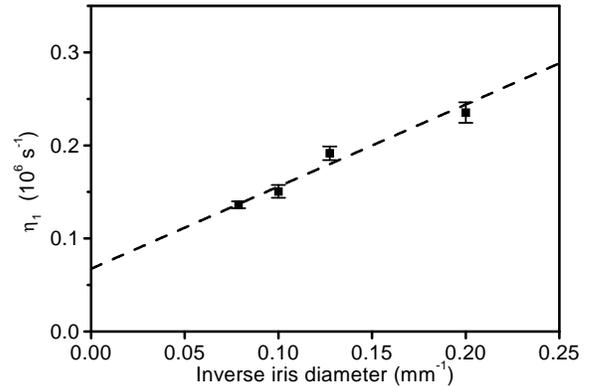,width=3.5in}}
\end{center}
\caption{Dependence of the decay rate $\protect\eta _{1}$ on light beam
diameter. Ligth intensity: $0.9\ mW/cm^{2}$.}
\label{irisdepend}
\end{figure}

Using the same intensity than in the case of Fig. \ref{intensdepeit}d, we
have checked that $\eta _{1}$ is linearly dependent on the inverse diameter
of the diaphragm placed before the cell indicating that this decay rate is
essentially determined by the atomic time-of-flight (see Fig. \ref
{irisdepend}).

\section{Theoretical analysis and discussion.}

The experimental results presented above will be discussed in this section
in view of a simple theoretical model of the atomic evolution. We follow the
standard density matrix approach using optical Bloch equations and the
rotating wave approximation \cite{LI95}. Several simplifications are made.
The atoms are considered at rest and the atomic sample is assumed to be
homogeneous. The finite time-of-flight of the atoms through the light beam
is taken into account in the calculation through a phenomenological decay
rate \cite{JYOTSNA95,ARIMONDO96}. The theoretical model does not intend to
represent the actual level structure of the $D_2$ transitions of $^{87}$Rb.
Instead, we have chosen to analyze two model transitions: $F_g=1\rightarrow
F_e=0$ and $F_g=1\rightarrow F_e=2$ which are the simplest to correspond to
EIT and EIA respectively \cite{LEZAMA99}. The two transitions are considered
closed in the sense that the radiative decay of the excited level is
exclusively into the ground level.

Following the procedure and notation introduced in \cite{LEZAMA99,LEZAMA00},
we consider an atom at rest with a ground level $g$ and an excited level $e$
with angular momenta $F_{g}$ and $F_{e}$ respectively and energy separation $%
{}\hbar \omega _{0}$. Spontaneous emission from $e$ to $g$ occurs at a rate $%
\Gamma $. The finite interaction time is accounted for by the relaxation
rate $\gamma $ ($\gamma \ll \Gamma $). The atoms are submitted to the action
of a magnetic field $B$ and a classical monochromatic electromagnetic field: 
$\vec{E}(t)=E\hat{e}\exp (i\omega t)$,$\ \ $where $\hat{e}$ is a complex
unit polarization vector.

Introducing the slowly varying matrix $\sigma =P_{g}\rho P_{g}+P_{e}\rho
P_{e}+P_{g}\rho P_{e}\exp \left( -i\omega t\right) +P_{e}\rho P_{g}\exp
\left( i\omega t\right) $ (where $\rho $ is the density matrix in the
Schr\"{o}dinger representation and $P_{g}\ $and $P_{e}\ $are projectors on
the ground and excited subspaces respectively), the time evolution of the
system (in the rotating wave approximation) obeys: 
\begin{eqnarray}
\frac{d\sigma }{dt} &=&-\frac{i}{\hbar }\left[ H_{Z}+\hbar \Delta
P_{e}+V,\sigma \right] -\frac{\Gamma }{2}\left\{ P_{e},\sigma \right\}
\label{bloch} \\
&&+\Gamma \left( 2F_{e}+1\right) \sum_{q=-1,0,1}Q_{ge}^{q}\sigma
Q_{eg}^{q}-\gamma \left( \sigma -\sigma _{0}\right)  \nonumber
\end{eqnarray}
where $H_{Z}=(\beta _{g}P_{g}+\beta _{e}P_{e})F_{z}B$ is the Zeeman
Hamiltonian ($\beta _{g}$ and $\beta _{e}$ are the ground and excited sate
gyromagnetic factors and $F_{z}$ is the total angular momentum operator
projection along the magnetic field); $\ Q_{ge}^{q}=Q_{eg}^{q\dagger }\ \
(q=-1,0,1)$ are the standard components of the vectorial operator defined
by: $\vec{Q}_{ge}=\vec{D}_{ge}\left\langle g\Vert \vec{D}\Vert
e\right\rangle ^{-1}$ where $\vec{D}_{ge}\equiv P_{g}\vec{D}P_{e}$ and $%
\left\langle g\Vert \vec{D}\Vert e\right\rangle \ $is the reduced matrix
element of the dipole operator between $g$ and $e$; $\Delta \equiv \omega
_{0}-\omega $ is the optical field detuning and $V=\left( \Omega /2\right)
\left( \hat{e}\cdot \vec{Q}_{ge}+\hat{e}^{\ast }\cdot \vec{Q}_{eg}\right) $
with $\Omega $ the reduced Rabi frequency: $\Omega =E\left\langle g\Vert 
\vec{D}\Vert e\right\rangle \hbar ^{-1}$. $\gamma \sigma _{0}$ represents a
constant pumping rate (due to the arrival of fresh atoms) in the isotropic
state $\sigma _{0}=P_{g}/\left( 2F_{g}+1\right) $. For a given solution of
Eq. \ref{bloch}, the instantaneous atomic absorption rate $w\left( t\right) $
can be evaluated using: 
\begin{equation}
w\left( t\right) \varpropto -iTrace\left[ \sigma _{eg}\left( \hat{e}\cdot 
\vec{D}_{ge}\right) -\sigma _{ge}\left( \hat{e}^{\ast }\cdot \vec{D}%
_{eg}\right) \right]  \label{instabs}
\end{equation}
where $\sigma _{eg}=\sigma _{ge}^{\dagger }=P_{e}\sigma \left( t\right)
P_{g} $.

Eq. \ref{bloch} represent a system of coupled first order linear
differential equations for the coefficients of $\sigma $. Using the
Liouville method the matrix elements of $\sigma $ can be organized into a
vector ${\bf y}$ and Eq. \ref{bloch} rewritten in the form: 
\begin{equation}
\frac{d{\bf y}}{dt}=M{\bf y}+{\bf p}_{0}  \label{Liouville}
\end{equation}
where $M$ $\left( e,\Delta ,\Omega ,\Gamma ,\gamma ,B\right) $ is a matrix
and ${\bf p}_{0}$ a constant vector corresponding to the pumping term $%
\gamma \sigma _{0}$.

\begin{figure}[tbp]
\begin{center}
\mbox{\epsfig{file=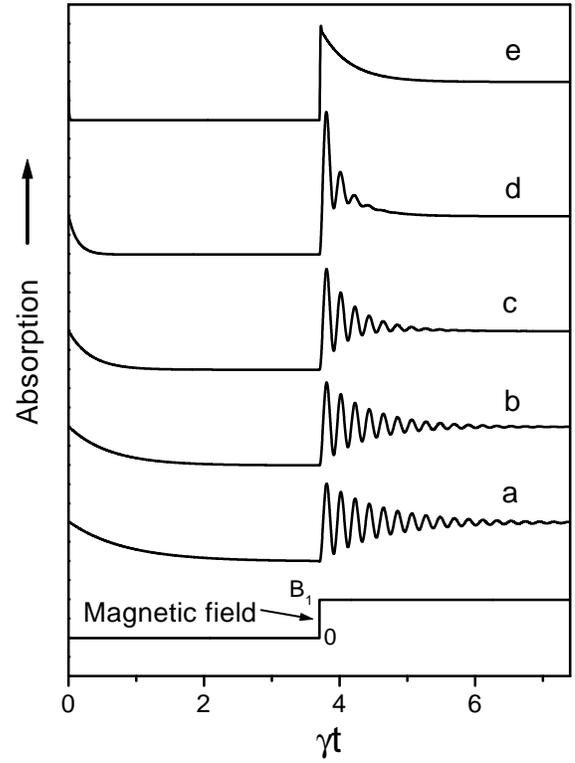,width=3.5in}}
\end{center}
\caption{Numerically simulated transients for the EIT type transition $%
F_{g}=1\rightarrow F_{e}=0$ for different optical field intensities. $\Omega
^{2}/\Gamma ^{2}=2\times 10^{-3}$ (a), \ $6\times 10^{-3}$ (b), $0.02$ (c), $%
0.06$ (d), $2$ (e) ($\Delta =0$, $\protect\gamma =0.002\Gamma $, $\protect%
\beta _{g}B_{1}=0.03\hbar \Gamma $).}
\label{simuleit}
\end{figure}

\begin{figure}[tbp]
\begin{center}
\mbox{\epsfig{file=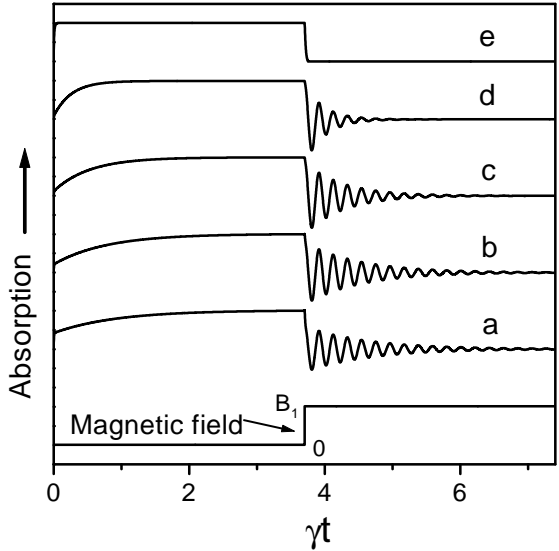,width=3.5in}}
\end{center}
\caption{Numerically simulated transients for the EIA type transition $%
F_{g}=1\rightarrow F_{e}=2$ for different optical field intensities. $\Omega
^{2}/\Gamma ^{2}=2\times 10^{-3}$ (a), \ $6\times 10^{-3}$ (b), $0.02$ (c), $%
0.06$ (d), $2$ (e) ($\Delta =0$, $\protect\gamma =0.002\Gamma $, $\protect%
\beta _{g}B_{1}=0.03\hbar \Gamma $).}
\label{simuleia}
\end{figure}

The solution ${\bf y}(t)$ of Eq. \ref{Liouville} was numerically calculated
for a magnetic field periodically alternating between two constant values $%
B_{0}=0$ and $B_{1}\neq 0$. The optical field was taken linearly polarized
in the direction perpendicular to the magnetic field. The results are
presented in Fig. \ref{simuleit} and \ref{simuleia} for different values of $%
\Omega /\Gamma $ and $\Delta =0$, $\gamma =0.002\Gamma $ and $\beta
_{g}B_{1}=0.03\hbar \Gamma $. For small values of $\Omega /\Gamma $ the $B=0$
and $B\neq 0$ transients show decay times of the order of $\gamma $ for both
transitions. The behavior is rather different at larger intensities where
the ($B\neq 0$) transient is of comparable duration to the $B=0$ transient
for the EIA type transition but is much slower for the EIT type transition.
In the latter case, the $B\neq 0$ transient clearly deviates from a simple
sine-damped evolution and approaches a pure exponential decay in the large
intensity limit with a characteristic rate of the order of $\gamma $.

A deeper insight into the transient evolution of these systems can be
obtained by the analysis of the eigenvalues $\lambda _{i}$ and eigenvectors%
{\bf \ }${\bf v}_{i}$ of matrix $M$ which can be numerically calculated ($%
i=1,\ldots ,n$ with $n=16$ and $n=64$ for the considered EIT and EIA
transitions respectively). All the real parts of the $\lambda _{i}$'s are
negative as expected for a stable system. For $B=0$ all the $\lambda _{i}$'s
are real indicating that the equilibrium will be reached through exponential
decays. For $B\neq 0$ some of the $\lambda _{i}$'s are complex indicating an
oscillating behavior as experimentally observed. For small values of $\Omega
/\Gamma $, the $\lambda _{i}$'s can be separated in three groups depending
on whether the corresponding absolute values of their real parts are of the
order of $\gamma $ or approaches $\Gamma /2$ \ or $\Gamma $. Group $1$ of
eigenvalues, which is the one of interest in this paper, is associated to
the evolution of the ground state coherences and populations, groups $2$ and 
$3$ are related to the relaxation of the optical coherence and excited state
populations respectively.

In general, the leading eigenvalue dominating the temporal evolution of the
atomic response should be the smallest (observable) one. Since the evolution
matrix $M$ includes the escape of the atoms from the interaction region (at
rate $\gamma $) it is quite obvious that no $\left| \lambda _i\right| $ can
be smaller than $\gamma $. As a matter of fact, a constant eigenvalue $%
\lambda _1\equiv -\gamma $ is always present. However, as will be discussed
next, the corresponding decay mode is unobservable. In any case, in the
limit of low driving field intensity, the leading eigenvalues approach the
time-of-flight decay constant $\gamma $ as expected.

The general solution of Eq. \ref{Liouville} is given by: 
\begin{equation}
{\bf y}(t)=\sum_{i}a_{i}{\bf v}_{i}exp(\lambda _{i}t)-M^{-1}{\bf p}_{0}
\label{gensolution}
\end{equation}
where the coefficients $a_{i}$ depend on the initial state of the system.
Two conditions are required on a given decay mode to be observable: $a)$ The
corresponding eigenvector ${\bf v}_{i}$ must be present in the decomposition
(\ref{gensolution}) of ${\bf y}(t)$ ({\it i.e.} $a_{i}\neq 0$). $b)$ The
density matrix $\sigma _{i}$ associated to eigenvector ${\bf v}_{i}$ should
correspond to a nonzero absorption of the incident optical field.

Using as initial conditions the steady state solutions of Eq. \ref{Liouville}
corresponding to $B=0$ or $B=B_{1}$, we have identified the observable decay
modes verifying condition $a)$ . The condition $b)$ can be checked by
calculating the absorption corresponding to $\sigma _{i}$ according to Eq. 
\ref{instabs}.

\begin{figure}[tbp]
\begin{center}
\mbox{\epsfig{file=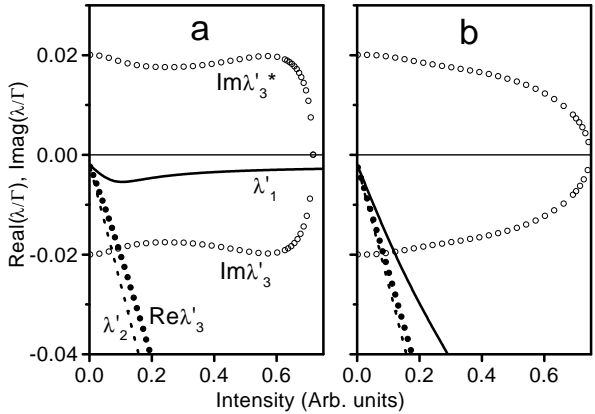,width=3.5in}}
\end{center}
\caption{Calculated values of the (group 1) observable eigenvalues of the
evolution matrix $M$ as a function of light intensity. a) EIT type
transition $F_{g}=1\rightarrow F_{e}=0$. b) EIA type transition $%
F_{g}=1\rightarrow F_{e}=2$. The dashed lines represent real eigenvalues
corresponding to the $B=0$ transients. Solid lines represent real
eigenvalues of the $B\neq 0$ \ transients. Circles represent complex
eigenvalues ( solid: real part, hollow: imaginary part) of the $B\neq 0$ \
transients. In (a) the intensity units correspond to the $\Omega ^{2}/\Gamma
^{2}$ ratio. An increase by a factor $2.5$ of $\left\langle g\Vert \vec{D}%
\Vert e\right\rangle ^{2}$ was assumed in (b).}
\label{eigeit}
\end{figure}

Figs. \ref{eigeit} shows the eigenvalues, corresponding to group $1$, of the
observable decay modes as a function of the optical field intensity for the
two transitions considered. The decay mode corresponding to $\lambda
_{1}\equiv -\gamma $ is unobservable since the corresponding decay is
exactly compensated by the pumping term ${\bf p}_{0}$ describing arrival of
fresh atoms.

Let us now discuss in more detail the $F_{g}=1\rightarrow F_{e}=0$
transition. This discussion can be simplified by noticing that this system
is totally equivalent to the open $\Lambda $ system formed by states $\left|
F_{g}=1,m=-1\right\rangle $, $\left| F_{e}=0,m=0\right\rangle $ and $\left|
F_{g}=1,m=1\right\rangle $ which can ``leak'' through spontaneous emission\
into the ``sink'' state $\left| F_{g}=0,m=0\right\rangle $. This open $%
\Lambda $ system has been studied in detail by Renzoni and coworkers \cite
{RENZONI99}. Following their steps, one can write optical Bloch equations
for the open $\Lambda $ system incorporating the time-of-flight relaxation
constant $\gamma $ for all levels. Analytical expressions of the eigenvalues 
$\lambda _{i}^{\prime }$ of the corresponding linear differential equations
system can be obtained for $B=0$ \ as a function of the relaxation rates,
the Rabi frequency $\Omega $ and the branching ratio $\alpha $ to the sink
state. One can check that the observable eigenvalue is in this case $\lambda
_{2}^{\prime }\simeq -\gamma +{\cal O}({\Omega }^{2}/\Gamma ^{2})$ in
agreement with Fig. \ref{eigeit}a. For $B\neq 0$ the $\lambda _{i}^{\prime }$%
's have to be evaluated numerically.

Taking $\alpha =1/3$, the open $\Lambda $ system exactly describes the $%
F_{g}=1\rightarrow F_{e}=0$. The use of the simpler $\Lambda $ system helps
to the identification of the eigenmode corresponding to a given $\lambda
_{i}^{\prime }$. For $B=0$ independently of the optical field intensity, the
smallest eigenvalue is $\lambda _{1}^{\prime }\equiv -\gamma $. The
corresponding eigenmode is, as expected, the dark state: $\left| \psi
_{D}\right\rangle =\sqrt{1/2}\left( \left| F_{g}=1,m=-1\right\rangle -\left|
F_{g}=1,m=1\right\rangle \right) $. Since the dark state is not coupled to
the light, it's transient temporal evolution is unobservable. This is not
longer the case for $B\neq 0$ since then the dark state is not stationary
and consequently the eigenmode corresponding to $\lambda _{1}^{\prime }\sim
-\gamma $ is contaminated with the bright state $\left| \psi
_{B}\right\rangle =\sqrt{1/2}\left( \left| F_{g}=1,m=-1\right\rangle +\left|
F_{g}=1,m=1\right\rangle \right) $ and thus coupled to the excited state. As
a consequence the transient evolution corresponding to $\lambda _{1}^{\prime
}\sim -\gamma $ becomes observable. It remains the leading eigenvalue even
at large values of the driving field intensity. This, together with the fact
that the oscillating mode decays at rate $%
\mathop{\rm Re}%
\lambda _{3}^{\prime }$ with $\left| 
\mathop{\rm Re}%
\lambda _{3}^{\prime }\right| >\left| \lambda _{1}^{\prime }\right| $ (see
Fig. \ref{eigeit}a) explains the slow exponential component of the $B\neq 0$
transient obtained for large optical field intensity in the EIT type
transition (Fig. \ref{simuleit}). The dependence on driving field intensity
of$\ \lambda _{1}^{\prime }$ presents a minimum at intensity $I_{c}$ ($%
\Omega ^{2}/\Gamma ^{2}\simeq 0.1$ in Fig. \ref{eigeit}a). Below this
critical value, the driving field intensity is responsible for the faster
damping of the transient. Above $I_{c}$ an increase in light intensity
results in the slowing of the atomic evolution.

The situation is rather different for the EIA type transition $%
F_{g}=1\rightarrow F_{e}=2$ (Fig. \ref{eigeit}b). In this case, all
observable decay modes correspond to eigenvalues whose real parts are (in
absolute value) increasing functions of the driving field intensity. As a
consequence, there is no exponential decay surviving significantly longer
than the damped oscillation for $B\neq 0$.

The simple model calculation presented above explains the main features of
the experimental observations. This is somehow surprising in view of the
several simplifications of the model with respect to the actual experimental
conditions. A first simplification is the neglecting of the atomic motion.
It is justified by the fact that the Raman resonance condition between
ground state Zeeman sublevels is unaffected by the Doppler effect. In
addition, the model does not account for the effect of the optical intensity
distribution in the beam profile and of the light propagation across the
sample. The influence of the former effect is minimized in the experiments
by only collecting light from the central (uniform intensity) portion of the
light beam. No significant variation of the transients with the atomic
density (optical thickness) of the sample was observed.

We discuss now the role of the contribution of the different hyperfine
transitions to the observed transients. The hyperfine structure of the
ground level of Rb is well separated in the absorption spectrum. However,
due to the Doppler broadening the excited state hyperfine structure is
unresolved. Consequently, for a given frequency position of the driving
field coupled to one of the ground state hyperfine levels $5S_{1/2}\left(
F_{g}\right) $, the atomic response is due to three different transitions to
excited state hyperfine levels $5P_{3/2}\left( F_{e}=F_{g},F_{g}\pm 1\right) 
$. The contribution of each hyperfine transition on the total absorption
signal is a function of the specific isotope and transition considered, the
precise position of the optical frequency within the Doppler absorption
profile and the light intensity. Nevertheless, it was shown that the total
absorption presents EIT type coherence resonances when the lower ground
state hyperfine level is excited and EIA type resonances for the case of the
upper ground state hyperfine level \cite{LEZAMA99,DANCHEVA00,RENZONI01}. In
the case of the lower ground state level, this is due to the fact that all
three hyperfine transitions give rise to EIT \cite{LEZAMA99}. In the case of
the upper ground state hyperfine level, the atomic response is
quantitatively dominated by the closed transition [$5S_{1/2}\left(
F_{g}=2\right) \rightarrow 5P_{3/2}\left( F_{e}=3\right) $ for $^{87}$Rb]
which corresponds to EIA. However this is only true provided that the
exciting laser is not too far detuned to the red side of the Doppler
absorption profile and that the Rabi frequency remains small in comparison
with the excited state hyperfine structure. In fact, significant distortions
in the transients were observed for the EIA type transition at maximum
available light intensity (not presented). Finally let us remind that for
simplicity we have theoretically analyzed the model EIA\ type transition $%
F_{g}=1\rightarrow F_{e}=2$. However, we have checked that qualitatively
similar results are obtained for the transition $F_{g}=2\rightarrow F_{e}=3$
occurring in $^{87}$Rb. The good agreement obtained between the observations
and the prediction of the simplified model, where a single atomic transition
is considered, suggests that the essential features of the atomic response
are a direct consequence of the type of the dominant transition(s) rather
than the specific transition(s) involved.

The most intriguing result presented above is the unexpected an rather
counterintuitive long transient observed for $B\neq 0$ with driving field
intensities near saturation for the EIT type resonances. Instead of being
the cause of the rapid damping of the atomic evolution, the applied driving
field is, in this case, responsible for slowing down the evolution. This is
the consequence of the Zeno effect\cite{ITANO90} in the sense recently
discussed by Luis\cite{LUIS01}. In this context, the optical field is seen
as a continuous measurement projecting the state of the system onto the DS
and thus preventing its evolution. Also, increasing the driving field
intensity results in an enhanced stability of the initial quantum state.
Following the analysis of Luis, the present result can be seen as the
preparation of a specific quantum state (the DS in our case) and its
preservation via the Zeno effect. Indeed, a transient such as the one
presented in Fig. \ref{simuleit}e correspond to a slow non-oscillating
evolution of the population in the DS. As Fig. \ref{eigeit}a indicates, the
survival time of the DS increases once the driving field intensity is above $%
I_{c}$. Although the quantum Zeno effect is usually presented in terms of
the projection postulate associated to measurement in quantum mechanics,
such picture is not essential in our case where the whole dynamics is well
reproduced by the Bloch equation treatment\cite{BEIGE96,LUIS01}. The
preservation of trapping states in optical pumping experiments at large
optical intensities was first reported in \cite{SLIJKHUIS86}. More recently,
a similar slowing down of the atomic evolution was discussed and observed by
Godun {\it et al. }\cite{GODUN99}.

From a similar perspective, the rather different temporal evolution observed
for Hanle/EIA resonances with $B\neq 0$, can be immediately explained by the
non existence of a state uncoupled to the light field onto which the system
could be projected. As a consequence, the observed and calculated decay
rates are increasing functions of the optical intensity.

\section{Conclusions.}

The transient evolution of the atomic absorption of a linearly polarized
optical field as the longitudinal magnetic field is suddenly switched on or
off has been observed in Rb vapor. Different transients have been observed
depending on the corresponding coherence (Hanle) resonance being of the EIT
or EIA type. The main features of the experimentally observed transients are
well reproduced by a theoretical model based on the numerical integration of
optical Bloch equations where the Zeeman degeneracy of the atomic levels is
fully taken into account. As expected, for low driving field intensities all
transient evolutions are governed by the time-of-flight relaxation rate. The
falling into the DS or the EAS ($B=0$ transients) occurring for EIT and EIA\
type transitions respectively, happens with similar decay rates that are
increasing functions of the driving field intensity. Interesting differences
arise between the $B\neq 0$ transients corresponding to the departure from
the DS or the EAS for driving field intensities approaching saturation.
While the EIA transient is rapidly shortened with the increase of the
driving field intensity, the EIT transient shows a slow non oscillating
component whose decay time is quite insensitive to the optical field
intensity. The latter result is interpreted as the preservation of the
initial quantum state via the Zeno effect.

\section{Acknowledgments.}

The authors are thankful to S. Barreiro for his collaboration in the initial
stages of the experiment and to J. Fernandez and A. Saez for technical
assistance. This work was supported by the Uruguayan agencies:\ CONICYT,
CSIC and PEDECIBA.


\end{document}